\magnification=1200
\baselineskip=6mm
\def\sect#1{\vskip 8mm \noi {\bf #1} \vskip 6mm}
\def\section#1{\vskip 8mm \noi {\bf #1} \vskip 6mm}
\def\subsection#1{\vskip 6mm \noi {\bf #1} \vskip 6mm}
\def\Ha{H$\alpha$}
\def\v{\vskip 2mm}

\def\noi{\noindent}

\def\title#1{\centerline {\bf #1}\vskip 8mm}
\def\author#1{\centerline{#1}\vskip 2mm}
\def\inst#1{\centerline{\it #1}\vskip 4mm}
\def\reference{\vskip 6mm{\noindent{\bf References}} \vskip 2mm}

\def\section#1{\vskip 8mm \noi {\bf #1} \vskip 8mm}

\def\r{\hangindent=1pc  \noindent}
\def\cen{\centerline}

\def\endpage{\vfil\break}

\def\kms{km s$^{-1}$}

\def\deg{$^\circ$}

\def\kms{km s$^{-1}$}
\def\Msun{M_{\odot \hskip-5.2pt \bullet}}

\def\deg{$^\circ$}

\def\ioa{Institute of Astronomy, University of Tokyo, Mitaka, Tokyo 181, Japan}
\def\email{sofue@mtk.ioa.s.u-tokyo.ac.jp}

\cen{\bf Nuclear-to-Outer Rotation Curves of Galaxies in the 
CO and HI lines}

\vskip 6mm
\author{Yoshiaki SOFUE}

\inst{\ioa}

\inst{E-mail: \email}

\centerline{(PASJ, Vol 49, No.1, 1997, in press)}
\vskip 8mm

\centerline{Abstract} 

We have derived rotation curves of nearby galaxies, which are 
almost completely sampled from the inner to outer regions. 
We used high-resolution position-velocity 
diagrams in the CO line along the major axes for the inner regions,
and HI-line data for the outer regions.
We combined the CO and HI data to obtain total rotation
curves from the nuclear to outer regions.
The rotation curves show a steep nuclear rise
within a radius smaller than the resolutions of CO-line 
observations, indicating a compact massive concentration 
near the nucleus.
We show that the nuclear steep rise is general and universal, 
and a rigid body-like gentle rise is exceptional. 

Keywords: Galaxies: kinematics -- Galaxies: rotation -- 
Galaxies: structure --  Galaxies: general  -- ISM: 
CO emission -- ISM: HI gas

\sect{1. Introduction}

Rotation curves (RC) of galaxies obtained  by optical (\Ha)
and  HI 21-cm line observations have been known to show a
rigid-body like increase in the central few kpc region, followed
by a flat rotation in the disk and outer regions 
(Rubin et al 1980, 1982; Bosma 1981b;  Persic et al 1995).
Since the HI gas is distributed widely in the galactic disk, 
even extending to the edges of galaxies, HI rotation curves are 
most useful for investigating the mass distribution in the outer region, 
and  have been used for studying the massive dark
halo (Kent 1987).
On the other hand, the HI gas is deficient in the central
region, where  the molecular hydrogen dominates (Sofue et al 1993).
CO-line observations have been obtained for a large number of
galaxies in the decade (Young and Scoville 1992), 
and position-velocity diagrams in the central region of these
galaxies have been obtained.
The CO rotation curves for the nuclear regions have suggested a much 
sharper rise in the central few hundred parsecs than has been known 
by HI or optical observations, which has been observed indeed for 
the Milky Way Galaxy (Clemens 1985).

Combining the CO-line rotation curves for the central regions with
HI and optical RC in the outer disks, we have obtained
'most completely sampled rotation curves' for several nearby galaxies
(Sofue 1996; Paper I).
The radio (CO + HI) rotation curves are found to have 
a sharp nuclear rise, often associated with a high-velocity peak,
 in the central few hundred parsecs, then a gap followed
by a second maximum at 5 to 10 kpc, and a flatter
or slowly declining outer part.

In this paper, we present radio (CO + HI), and optical rotation 
curves for a larger number of spiral galaxies using 
high-resolution CO and HI-line data.
We aim at clarifying if the nuclear rise of rotation curves
is indeed a universal and general characteristic of rotation
of spiral galaxies.

\sect{2. Envelope-Tracing Method for Deriving Rotation Curves}

We adopt the envelope-tracing method as used in Paper I to 
derive rotation curves, which  uses the loci of terminal  
velocity in position-velocity (PV) diagrams.
We define the terminal velocity by a velocity at which
the intensity becomes equal to
$$
I_{\rm t}=[(0.2 I_{\rm max})^2+I_{\rm lc}^2]^{1/2} \eqno(1)
$$
on the PV diagrams, where $I_{\rm max}$ and $I_{\rm lc}$ are 
the maximum intensity and intensity corresponding to the lowest 
contour level, respectively.
This equation defines a 20\% level of the intensity profile at a fixed 
position, $I_{\rm t}\simeq 0.2 \times I_{\rm max}$,
if the signal-to-noise ratio is sufficiently high.
If the intensity is not high enough, the equation gives
$I_{\rm t}\simeq I_{\rm lc}$, which approximately defines the loci 
along the lowest contour level ($\sim 3 \times$ rms noise).
The terminal velocity is then corrected for the 
velocity dispersion of the interstellar gas
($\sigma_{\rm ISM}$) and the velocity resolution of observations 
($\sigma_{\rm obs}$) as 
$$
V_{\rm t}^0=V_{\rm t}-(\sigma_{\rm obs}^2 + \sigma_{\rm ISM}^2)^{1/2}.
\eqno(2)
$$ 
Small-scale structures due 
to clumpy ISM and clouds, and partly due to the noise
in the observations, are smoothed by eye estimates.
Since it was not practical to apply these procedures automatically
by a computer, we have drawn the curves by hand on each 
position-velocity diagram.
The rotation velocity is finally obtained by 
$$ 
V_{\rm rot}=V_{\rm t}^0/{\rm sin}~i, \eqno(3)
$$
where $i$ is the inclination angle of the disk plane.
The accuracy of determining the terminal velocity, and therefore, 
the accuracy of the obtained rotation curve was typically 
$\pm 10 \sim 15 $/sin $i$ \kms.

Simply-traced envelopes on 
the two sides of the nucleus have a discontinuity at the nucleus
due to the finite beam width.
We have avoided this discontinuity by stopping the tracing
at a radius corresponding to the telescope resolution, and then by 
connecting the both sides of rotation curve by a straight
(solid-body like) line crossing the nucleus at zero velocity. 
Figure 1 shows examples of the CO-line position-velocity
diagrams and traced rotation curves for several galaxies, 
combined with HI rotation curves.
The pair of rotation curves thus obtained on both (receding
and approaching) sides of the nucleus are finally averaged to 
give a rotation curve as a function of the radius.
The inner rotation curves thus derived from CO-line data
were, then, connected smoothly to the HI curves in the outer disk.
In most cases the discrepancy between CO and HI rotation velocities
at the connecting area were within about $\pm 10$ kms.
Since the CO data usually have higher resolution than HI, we adopted
CO curves, when both CO and HI curves were available at the same radius.
 
\subsection{3. CO + HI Rotation Curves for Individual Galaxies}

Figure 2 shows the obtained radio (CO+HI) rotation curves for the
studied galaxies with the basic data inserted, such as the inclination,
position angle of the major axis, and a possible distance
taken from the literature cited in table 1.
These distances are often uncertain, and therefore, the linear scales indicated
along the upper ordinates are also uncertain.
Vertical bars of the crosses represent the typical errors in  
velocity determination, and the horizontal bars indicate the angular 
resolutions of observations.
Vertical dashed lines separate each rotation curve into two or
three parts, where the CO and HI data have been used, respectively.
When different data with different angular resolutions are combined,
their border is also indicated by a dashed line.
We show by inset optical images of the galaxies taken from the
STScI Digitized Sky Survey (DSS).
A $10'\times10'$ field is displayed for each object
(larger fields for a few galaxies), and the angular
scale is given along the major axis.
We describe the individual galaxies below.
Parameters and references for the data are listed in Table 1.
 
---- Fig. 1 and 2; Table 1 ----

NGC 224 (M31): 
This galaxy is one of the few exceptional nearby galaxies of Sb type,
which emit very weak CO line in the center: 
Even in our deep observation with the Nobeyama 45-m telescope
we could detect only weak or almost no CO emission in the nuclear 
region (Sofue and Yoshida 1993).
Therefore, no sufficient kinematical data are available to derive 
a CO rotation curve.
However, there have been a number of studies of the rotation 
curve as summarized in Sofue and Kato (1981). We revisited and 
obtained a new RC using the optical data in the [NII] 
$\lambda 6583$ emission observed by Rubin and Ford (1970)
and in the H$\alpha$ line by Ciardullo et al. (1988). 
We adopted the H$\alpha$ velocities for the very inner region 
within $1'$, and those from [NII] line for the inner $6'$, 
where the rotation rises steeply to a peak velocity of 
220 \kms\ at $2'$ radius.
Beyond this radius, we adopted the HI data from the Bonn 100-m 
telescope, which is sensitive enough to trace the rotation with a 
sufficient accuracy, although the resolution ($9'$) was not
enough to resolve the innermost region. 
The HI position-velocity diagram shows a pair of symmetrical
shoulders at $\pm10'$ with respect to the nucleus, corresponding to 
the central peak indicated by the optical observations.
The velocity, then, increases to a maximum as high as 280 \kms\
at $40'$, and declines slowly toward the edge.

NGC 253: This is a CO-rich starburst galaxy, and has been studied in 
Paper I based on a lower-resolution CO data. We revisited this galaxy
using the higher-resolution CO data from the 45-m telescope. 
The central peak of rotation curve is more clearly 
seen in the new data. See Paper I.

IC 342: A central rise and a step, corresponding to the nuclear
peak at 130 \kms, is followed by an almost flat rotation in the
disk and outer part. See Paper I. 

NGC 598 (M33): Because of its proximity, we used lower-resolution CO data
from the Kitt Peak 12-m telescope, which we combine with the HI
rotation curve. 
The CO rotation observed for the inner $5'$ region
agrees with the HI rotation.
The rotation velocity increases 
steeply near the nucleus, but reaches only to 20 \kms, 
and turns to increase more gently  from the inner $1'$ region to $10'$. 
It continues to increase monotonically to the observed edge at $30'$.
This galaxy shows neither a steep rise 
nor a sharp central peak, similarly to NGC 4631, and is exceptional
among the galaxies studied here.
Note that both M33 and NGC 4631 are low-mass galaxies. 

NGC 660: This is a nearly edge-on Sc galaxy with a polar ring 
(van Driel et al. 1995). We applied the envelope-tracing method
to the HI position-velocity
diagrams for the disk and polar ring, as well as to the CO (2-1)
position velocity diagram for the inner part.
The disk rotation curve can be apparently connected smoothly to that
of the polar ring.
Since the polar-ring rotation will manifest the gravitational potential 
in the halo, which is supposed to be round, we may approximate the
outer rotation curve in the disk plane by this polar ring curve. 
The CO line profiles 
indicate that the maximum rotation velocity occurs near the center,
in an unresolved region of $<10''$. 
The inner CO rotation has a high central peak of 
about 190 \kms, followed by a disk rotation, where HI and CO 
velocities coincide well.

NGC 1068: This is a Seyfert galaxy. The position angle of the
major axis is not well defined. We adopted
70\deg\ for the outer and disk region 
as read from the outer optical velocity field
as well as from the optical image,
and 90\deg\ for the central region from the CO velocity field
(see the literature cited in Table 1). 
The rotation velocity apppears to rise steeply in the center toward the peak on
the  molecular ring. However,  the very central behavior is not 
observed due to the lack in the CO emission at the center.

NGC 1097: If the adopted inclination angle of 40\deg\ 
is correct, the apparent rotation velocity of the molecular ring 
is as high as 350 \kms, showing a nuclear  peak.
The rotation decreases then to a dip at 40$''$ radius, followed
by a broad maximum of about 300 \kms\ at 2$'$, 
and then flattens and slowly declines toward the edge.
The general characteristic is similar to that of the Milky Way,
except the velocity amplitude.

NGC 1365: This is a typical barred spiral with a Seyfert nucleus.
The CO data from the SEST 15-m telescope indicate a steep rise and a
central peak at 240 \kms, followed by a disk component as high as
270 \kms, and then a monotonically declining rotation observed in the
HI line.

NGC 1808: A steep nuclear rise and a peak are followed by a gradual 
decrease toward the edge of the observed rotation curve. See Paper I.

NGC 2403: Due to its proximity, we used the lower-resolution CO data from
the Kitt Peak 12-m telescope. No higher-resolution
observations are available. 
The RC increases steeply in the center till 80 \kms, then
 increases slowly till $5'$ radius, where the curve turns to
be flat at a velocity as low as 120 \kms.
The CO and HI curves agree with each other at $1'$ to $5'$, while
CO velocity is much higher in the central 1$'$ region.

NGC 2841: Since no high-resolution CO data are available, we 
used the FCRAO 14m CO data for the inner 4$'$ to derive a CO
rotation curve, and combined it  with the outer HI rotation curve.
Although the very central kinematics is not clear due to the weak 
CO emission near the nucleus, a steep rise within the central 1$'$
region and a peak as high as 310 \kms are visible.
After increasing to a maximum of 330 \kms\ at 2 to 3$'$,
the velocity decreases and turns to be flat at 270 \kms until 
the edge.

NGC 2903: The CO rotation curve has a peculiar step at 1 -- 2$'$ radius
with a maximum of about 170 \kms\ at 1$'$. 
It, then, increases  to a second maximum of 200 \kms\ at 3$'$.
The step is, however, asymmetric with respect to the center:
This inner maximum has a higher velocity in the SW side than
in the NE.
The outer rotation from HI is, however, very flat until the edge.

NGC 3031 (M81): Optical and CO-line data have been combined for the
central $2'$. The rotation has a central peak at 
300 \kms, although  the peak is not very sharp. 
The HI rotation from the disk to outer region is similar to that
of the Milky Way in shape and velocity.

NGC 3034 (M82): This is an interacting peculiar galaxy with a 
starburst activity and is rich in CO gas.
The CO ($J=2$--1) data obtained by  the IRAM 30-m telescope
show a clearly declining rotation curve, 
well fitted by a Keplerian rotation.
The VLA HI data   agree with the CO data, 
though the error is larger.

NGC 3079: Sharp and high-velocity nuclear rise and a peak, and then
a gap,  are followed by a disk component, similar to that of the 
Milky Way.  The outer rotation is declining. See Paper I.

NGC 3198: The central steep rise stops at a peak as low as  50 \kms.
Then, the rotation increases in a rigid-body fashion, showing a next
step. The disk part has a broad maximum of 170 \kms,
followed by a flat rotation.

NGC 3521: The inner rotation curve increases steeply, and attains
a small and sharp central peak at 10$''$  of 210 \kms, 
followed by a dip at 20$''$. Then, it increases slowly till
a broad maximum at $1'.5$. 
The outer HI RC gradually decreases till the edge.

NGC 3628: The CO PV diagram as obtained by interferometer
observations with the Nobeyama Millimeter Array shows
a nuclear disk asymmetric with 
respect to the position-velocity center of the outer disk. 
We took the center of the nuclear disk
as the origin of the rotation curve, and connected the inner RC with the outer 
rotation curve by averaging the position and velocities in
both sides of the nucleus.
The CO+HI RC has a steep rise near the center within 5$''$,
indicating a nuclear disk, rotating at 190 \kms.
It further increases to 200 \kms\ at 1$'$, and is followed
by a flat rotation in the outer disk.

NGC 4258: The velocity increases to a maximum of 230 \kms within
the central 30$''$, and decreases to a flat
part at 3 to 7$'$. The outer HI rotation is almost flat, after a
broad maximum at 10$'$ radius.

NGC 4303: 
The rotation curve for the central 15$''$
has been derived using a position-velocity
diagram at a constant declination obtained by the NMA.
The correction for position-angle and inclination 
resulted in a high velocity peak, which is not clearly seen 
in the 45-m data. The error for the central peak is, therefore, 
as large as $\sim30$ \kms.
The disk and outer parts of the rotation curve are flat.

NGC 4321 (M100): The NMA observation
shows a steeply increasing rotation  within a 
$5''$  radius to a central peak of 230 \kms\ at 10$''$, 
followed by a dip at 20$''$.
The RC, then, increases gradually till 3$'$ to a broad maximum
of 270 \kms\ in HI velocity.
The outermost rotation velocity is not certain because of the
warped HI disk and low inclination.

NGC 4565: This is a typical edge-on galaxy of Sc type. The rotation 
has a central peak and a flat disk-to-outer rotation.
It has a sign of declining in the outermost region.
See Paper I.

NGC 4569: This is Virgo galaxy which shows a truncated HI gas disk.
CO-line position-velocity diagrams obtained with 
the NMA at a $9''$ resolution show a 
sharp rise to a 200 \kms\ central peak.
The 45-m data show a rotation minimum at $40''$, followed
by an increase toward the HI high-velocity rotation in the
outer disk. The outskirts of HI gas are missing due possibly to 
a ram-pressure stripping by the intra-cluster gas. 

NGC 4631: This is an interacting, edge-on semi-dwarf galaxy. 
The CO RC in the central 40$''$ is rigid-body like, having
neither a nuclear rise nor a peak.
The RC is followed by a flat and declining HI rotation in 
the outer disk.
This galaxy is one of the two exceptions (with M33), 
which show no nuclear rise of rotation in CO.
However, the intensity distribution along the major axis indicates
no concentration of CO gas in the center, but either 
a ring or clumps avoiding the nucleus (Sofue et al. 1989). 
Therefore, the apparent rigid-body rotation in the position-velocity
diagram may be due to the lack of gas near the nucleus, 
even if the true rotation curve has a sharper rise. 

NGC 4736: The rotation rises steeply within 12$''$, and
attains a central peak at 18$''$. 
It, then, decreases monotonically toward the observed edge.

NGC 4945: We used position-velocity diagrams along the major axis
from the two-elements interferometric observations
to derive an HI rotation curve,  and those from the SEST 15-m 
observations in the CO ($J=1$--0) and CO ($J=2$--1) emissions.
We adopted the higher-resolution CO (2--1) data for the inner region, 
CO (1--0) for the disk, and HI  for the outer region.
They agree in the overlapped regions, except that the CO (2--1)
data indicated a shaper central peak of RC than CO (1--0).
The obtained RC is similar to that of our Galaxy.
It has a steep rise and a sharp central peak,  followed by a 
broad maximum in the disk and a flat part in the outskirts.

NGC 5033: A sharp rise to a central peak at
10$''$ is followed by a second peak at 35$''$.
The outer HI rotation is flat till the edge, indicating a sign
of declining rotation near the edge.

NGC 5055: A sharp, rigid body-like rise within the central 30$''$
is followed by a flat part till a broad maximum at
4$'$. The rotation declines then till 8$'$, beyond which it
is nearly flat.

NGC 5194 (M51): A steep rise in the center is followed by a flat disk
rotation till $2'.7$ radius, superposed by high-amplitude 
velocity deviation due to the spiral arms. 
The rotation velocity, then, declines
more rapidly than a Keplerian curve.
This may be due to the gravitational disturbance by the companion 
NGC 5195, or to a warp in the outer disk likely caused by 
the tidal interaction. See Paper I.

NGC 5236 (M83): The RC rises very steeply near the center,
attaining a sharp peak at 10$''$ of 250 \kms. 
It, then, decreases steeply toward a deep dip at 30$''$
with a minimum velocity as low as 120 \kms.
The RC again increases toward a second broad maximum at 2$'$
of 190 \kms, and is followed by a gradually declining HI 
part in the outer disk.
The sharp central peak and deep dip are peculiar among the
galaxies so far known of the rotation curves, and may be
related to the bar and accretion process.

NGC 5457 (M101): This is a nearly face-on Sc galaxy, but no 
high-resolution CO data are available. Because of its 
large apparent size, we used lower-resolution CO data from the
Kitt Peak 12-m telescope.
The central rise within the beam width 30$''$ is recognized,
but the resolution is not sufficient for discussing the detail about
the nuclear rise. 

NGC 5907: This is an almost perfectly edge-on galaxy. 
The central RC rises steeply to a shoulder of 200 \kms\ at 10$''$.
The RC further continues to increase till 4$'$ radius, and, 
then, declines toward  the edge. See Paper I.

NGC 6946: The central peak is very sharp and as high as 240 \kms,
followed by a gap which is rather flat till the second increase 
toward  the disk maximum starts. 
This rotation curve is similar to that
of the Milky Way. See Paper I.
 
NGC 6951: The central high-velocity peak of 260 \kms\ is remarkable
at a radius of 20$''$. Since the resolution of the 45-m telescope
is not enough to resolve this region due to the relatively large
distance, the true peak velocity may be much higher.
The CO RC has been combined with the H$\alpha$ data for the
disk part, showing a gradually
increasing toward the outer region. 

NGC 7331: The central steep rise has a shoulder-like step of 
130 \kms\ at 14$''$, and reaches a peak of 240 \kms\ at 34$''$
after a rigid-body like increase.
The outer HI rotation  is almost flat, slightly
increasing  till the end of the disk.

Milky Way: For a comparison, we show the rotation curve for
our Galaxy. The Sun's distance from the nucleus is taken to be
8 kpc. The RC within the solar circle was taken from 
Clemens (1995), and that for the outer region from Honma and Sofue
(1996a, b). See the literature for details.

\sect{4. Discussion}

\subsection{4.1. Steep Nuclear Rise}

We have derived the CO rotation curves of the central regions for nearby
spiral galaxies, and combined them with  outer HI rotation curves. 
A remarkable feature obtained in the present study is the steep nuclear
rise of rotation within a radius smaller than the beam width of CO
observations. 
Two exceptional cases were found:
A nearby Sc galaxy M33 known for the low surface brightness
shows a rapid rise in the center, but to a low velocity, and then
a gentle rise;  
NGC 4631, a semi-dwarf edge-on galaxy of amorphous type,
has a rigid-body rise. However, NGC 4631's case 
might be an apparent phenomenon due to the lack of CO gas near the 
center, and the true rotation curve may have a sharper rise. 
Hence, the only one exception having indeed a slowly rising 
rotation is M33.
 
\subsection{4.2. The Nuclear Mass Component}

In Paper I we have shown that the steeply rising rotation curves 
can be generally fitted by a model with four mass components: 
the nuclear compact mass, central bulge, disk, and the massive halo.
Particularly, such a steep rise within the central few hundred 
parsecs as observed for NGC 3079, NGC 4321 and NGC 6946 
by high-resolution interferometer observations indicates the 
existence of a compact nuclear mass of a 
100 to 150 pc radius and  a mass of several $10^9\Msun$.
Hence, some of the galaxies are likely to be nested by a more compact
stellar mass concentration in the center than the usual bulges, 
which we called a bulge-in-bulge.

In order to avoid the discontinuity of envelope-traced terminal
velocities from both sides of the nucleus on the position-velocity 
diagrams,  we have connected them by a solid-body like line 
crossing the  nucleus at zero velocity within a radius comparable 
to the angular resolution.
Therefore, the   curves drawn within the
regions smaller than the angular resolution may not represent the
true velocities.
It is more likely that the true rotation curve has a
sharper, unresolved rise. It is also likely that the velocity  
remains finite till the nucleus, or even increases, 
since the galactic nuclei are often nested by a compact 
massive object.
 
\subsection{4.3. Circular Rotation-vs-Bar Debate}

By definition, a rotation curve is the trace of terminal
velocities in the position-velocity diagram along the major axis.
It may be superposed by non-circular motions such as due
to the density waves, bars and oval potential, or interaction
with the companion.
It is, therefore, not straightforward to derive the mass distribution
using the curves by assuming a circular rotation.
Nevertheless, this assumption has been extensively adopted 
in deriving the mass of galaxies including hypothetical dark halos. 
The  mass and potential derived from the circular-rotation assumption 
has also been used to discuss the inner and outer Lindblad resonances
or to analyze the pattern speed and density waves.
Rotation curves have also been used to derive the mass distributions
in the bulges and central cores, and have given a good
agreement with the surface photometry (e.g., Kent 1987).
These studies have made available the most important parameters
of the galaxies, and the circular-rotation assumption appears
reasonable in so far as the basic structures of galaxies are analyzed. 

On the other hand, it is a trend to introduce a bar in the analysis.
However, if one stands on a bar hypothesis,
the so-called rotation curves will give no more any quantitative 
information about the mass, unless the true potential is 
determined by independent observations. 
One might argue that the nuclear rise and peak of
rotation may be due to a high-velocity flow of 
gas along a bar parallel to the line of sight,
whereas the mass (potential) distribution is not known, 
which is necessary to calculate the flow.
However, the probability of looking at a bar from its end 
is far smaller compared to that of looking at it at a finite angle.
In the latter case, the gas  along the bar will be observed roughly at 
the bar's pattern speed, and, therefore, will show a rigid-body like
rotation at lower velocities than a circular rotation.
Hence, if some cases of the nuclear rise of rotation are due to 
a bar indeed, a larger number of galaxies must be observed to show a 
gentle rise in  a rigid-body fashion. 
This is, however,  not the case at all.
Note that M33 has no bar, and NGC 4631 is an amorphous dwarf and does 
not provide a firm case for rigid body rotation.  
Hence, it is not likely that the nuclear rise 
observed in almost all galaxies studied here is due to a bar-induced 
gas flow.

We may, therefore, reasonably assume that the central
rotation is circular in the zero-th order approximation, 
and the observed rotation curves 
manifest the mass distribution within the bulges.
This can be clarified by comparing the observed rotation
curves with those calculated by a mass model based on
a surface photometry  assuming an appropriate mass-to-light ratio, 
as has been obtained for disk regions by Kent (1987) on the basis of 
optical data. Since the nuclear region is highly obscured
in optical wavelengths, we may need infrared surface
photometry for the comparison, which will be described in a 
separate paper.


\reference

\r Ables, J. G., et al. 1987 MNRAS 226, 157

\r Begeman, K. G. 1989   A\&A 223, 47 

\r Bosma, A., van der Hulst, M., Sullivan, III. , W. T. 1977 

\r Bosma, A., Goss, W. M., Allen, R. J. 1981 A\&S 93, 106. 

\r Bosma, A. 1981a, AJ 86, 1791.  

\r Bosma, A. 1981b, AJ, 86, 1825. 

\r Bosma, A., van der Hulst, J. M., Sulivan, III, W.T. 1977, A\&A 57, 373.

\r Clemens, D. P. 1985 ApJ 295, 422. 

\r Casertano, S. 1983, MNRAS, 203, 735. 

\r Casertano, S., van Gorkom, J. H. 1991, AJ 101, 1231. 

\r Ciardullo, R., Rubin, V.C., Jacoby, G. H., 
Ford, H. C., and Ford, Jr. W. K. 1988 AJ 95, 438.

\r Combes F 1992 ARA\&A, 29, 195. 

\r Combes, F., Gottesman, S. T., Weliachew, L. 1977 A\&A 59, 181

\r Cram, T. R., Roberts, M. S., and Whitehurst, R. N. 1980 
	A\&AS 40, 215. 

\r Dahlem, M., Golla, G., Whiteoak, J. B., Wielebinski, R., 
H\"uttemessiter, S., and Henkel, C. 1993 A\&A 270, 29
 
\r Gerin, M., Nakai, N., Combes, F. 1988, A\&A 203, 44

\r Goad, J. W. 1976 ApJS 32, 89 

\r Guhathakurta, P., and van Gorkom, J. H. 1988 AJ 95, 851.
 
\r Honma, M., and Sofue, Y. 1996a PASJ, submitted

\r Honma, M., and Sofue, Y. 1996b PASJ Letter, in press.

\r Hunter, D. A., Rubin, V.C., R. and Gallagher, J. S., III 
	1986 AJ 91, 1086 

\r  Irwin, J.A. and Sofue, Y. 1996, ApJ 464, 738 

\r Kaneko, N., Satoh, T., Toyama, K., Sasaki, M., Nishimurfa, M., Yamamoto, M.
1992 AJ 103, 422 

\r Kaneko, N., Morita, K., Fukui, Y., Takahashi, N., Sugitani, K., 
 Nakai, N. Morita, K. 
1992 PASJ 44, 341.

\r Kaneko, N., Morita, K., Fukui, Y., Sugitani, K., Iwata, T., Nakai, N.
Kaifu, N., Liszt, H. 
1989 ApJ 337, 691. 

\r Kenney, J., Young, S. J. 1988 ApJS 66, 261.	

\r Kenney, J. D. P., Scoville, N. Z., and Wilson, C. D. 
1991 ApJ 366, 432.
   
\r Kent, S. M. 1986 AJ 91, 1301. 	

\r Kent, S. M. 1987 AJ 93, 816. 	

\r Knapen, J. H., Cepa, J., Beckman, J.E., Soledar del Rio, M., 
Pedlar, A. 1993 ApJ 416, 563 	

\r Knapen, J. H., Beckman, J. E., Heller, C. H., Shlosman, I. 
	De Jong R. S. 1995 ApJ, 454, 623

\r Newton,  K. 1980 MNRAS 190, 689 

\r Lake, G. and Feinswog, L. 1989 AJ 98 166 

\r Marguez, I., Moles, M. 1993, AJ 105, 2050. 

\r Nishiyama, K. 1995 PhD Thesis, University of Tokyo 

\r Ondrechen, M. P., van der Hulst, J. M. 1989 ApJ 32, 29.

\r Ondrechen, M. P., van der Hulst, J. M., Hummel, E.
	1989 ApJ 342, 39O  	

\r Planesas, P., Scoville, N., Myers, S. T. 1991 ApJ 369, 364.

\r Persic, S., Salucci, P., Stel, F. 1996 MNRAS.281, 27.

\r Rots, A. H. 1975 A\&A 45, 43 

\r Rubin, V. C., and Ford, Jr. W. K. 1970 ApJ 159, 381
 
\r Rubin, V. C., Ford, W. K., Thonnard, N. 1980, ApJ, 238, 471 

\r Rubin, V. C., Ford, W. K., Thonnard, N. 1982, ApJ, 261, 439 

\r Sage, L. I., and Westpfahl, D. J. 1991 A\&A 242, 371 

\r Sakamoto, K. 1995 PhD Thesis, University of Tokyo 

\r Sakamoto, K.,  Okumura, S., Minezaki, T., Kobayashi, Y., Wada, K. 
	1995 AJ 110, 2075 

\r Sandqvist, Aa., J\"ors\"aster, S., and Lindblad, P.O. 
	1995 A\&A 295, 585. 	


\r Sofue, Y. 1996, ApJ, 458, 120 (Paper I)

\r Sofue, Y., Doi, M., Krause, M., Nakai, N., and Handa, T. 
	1989 PASJ  41, 113 	

\r Sofue, Y., Handa, T., and Nakai, N. 1989 PASJ  41, 937. 

\r Sofue, Y., Handa, T., G. Golla, M. Krause, and R. Wielebinski
	1990, PASJ 42, 745

\r Sofue, Y., Honma, M., Arimoto, N. 1994, A\&A 296, 33-44. 

\r Sofue, Y. and Kato, R. 1981 PASJ 

\r Sofue, Y., Reuter, H.-P., Krause, M., Wielebinski, R., 
and Nakai, N. 1992 ApJL 395, 126.

\r Sofue, Y., Tutui, Y., Honma, M. 1997 in preparation

\r Sofue, Y., and  Yoshida, Y. 1993, ApJL 417, L63  
	 
\r Thornley, M. D., and Wilson,  C. D. 1995 ApJ 447, 616.

\r van Albada, G.D. and Shane, W.W. 1976 A\&A 42, 433. 	
 
\r van Driel, W. et al 1995 AJ 109, 942 
 
\r Weliachev, L., Sancisi, R., Gu{\'e}lin, M. 1978 A\&A 65, 37.

\r Wilding, T., Alexander, P., and Green, D. A. 1993 MN 263, 1075

\r Wilson, C. D., Scoville, N. 1989 ApJ 347, 743. 

\r Young, J. S. , Scoville, N. Z. 1992, ARA\&A, 29, 581.
 
\r Young, J.S., et al. 1995 ApJS 98, 219

\r Yun, M. S., Ho, P.T.P, and Lo, K.Y. 1993 ApJL 411, L17.

\endpage

\settabs 9 \columns

\def\v{\vskip 4mm}

\noindent Table 1: Parameters for galaxies and references for PV 
diagrams and rotation curves.
\vskip 2mm \hrule \vskip 2mm
\+ Galaxy & Type & Incl. & Dist.$^*$ 
& Line & Telescope$^\dagger$ & A. Reso.& References \cr
\+ & &(deg) & (Mpc) & & & (a.sec) & \cr
\vskip 2mm \hrule \vskip 2mm

\+NGC 224  (M31) Sb   77 &&&0.69 & Opt.[NII] && &Rubin \& Ford 1970\cr
\+ &&&&H$\alpha$ && &Ciardullo et al. 1988\cr
\+&&&&HI & E100m& 540 & Cram et al. 1980\cr

\v


\+ NGC 253  Sc && 78.5 &2.5&\dotfill&\dotfill&\dotfill&Paper I\cr
\+  & &  &  & CO &N45m &15 & Nishiyama 1995 \cr
 
\v

\+ IC 342 Sc &&25    &3.9  & \dotfill&\dotfill&\dotfill&Paper I\cr
\v

\+ NGC 593 (M33) Sc  54 &&& & HI & CI & $47\time93$ & Newton 1980 \cr
\+ &&& 0.79 & CO & K12m & 55 & Wilson \& Scoville 1989 \cr
\v

\+NGC 660 Sc&& 70& 13 & HI & W &20 &van Driel et al. 1995 \cr
\+&Polar ring &&& HI& W & 60 & ibid \cr
\+&&&& CO 2-1 & I30m & 12 & ibid \cr
 
\v
\+ NGC 891 Sb &&88.3  &8.9  &\dotfill&\dotfill&\dotfill&Paper I\cr
\v

\+NGC 1068 Sb Sy&& 46 &18.1 &Opt. &&& Kaneko et al. 1992\cr
\+ &&&& Opt && & Galleta \&Recillas-Cruz 1982 \cr
\+ &&&& CO & N45m & 17& Kaneko et al 1991 \cr
\+ &&&& CO & NMA  &$5\times 4$ &Kaneko et al 1992 \cr
\v

\+NGC 1097 SBb && 40& 16 & CO & N45m & 15& Gerin et al. 1988\cr
\+ &&&& HI & V & 25 & Ondrechen et al. 1989 \cr

\v

\+ NGC 1365  SBb &&46 & 15.6 & HI &V & $36\times23$ & Ondrechen \& van 
der Hulst 1989 \cr
\+& & &  & CO 2-1&S15m&25&Sandqvist et al. 1995\cr  
\v

\+ NGC 1808 Sbc & &58   &11.4 &\dotfill&\dotfill&\dotfill&Paper I\cr 
\v

\+NGC 2403 Sc &&60&3.25& HI &W&40/60& Lake \& Feinswog 1989\cr
\+ &&&&		CO &K12m &60& Thornley \& Wilson 1995\cr   

\v


\+NGC 2841 Sb& & 68& 9 & HI & W & $51\times65$ Bosma 1981a \cr
\+ &&&& CO &F14m & 45 & Young et al. 1995 \cr 

\v
\+NGC 2903 Sc &&35& 6.1 & HI& W &&Lake \& Feinswog 1989; Kent 1987\cr
\+ &&&& CO&N45m &15& Sofue et al 1997\cr
\v

\+ NGC 3031  (M81)  Sb 59 &&& 3.25 & Opt & & &Goad 1976 \cr
\+ &&&& CO & K12m & 55& Sage \& Westpfahl 1991 \cr
\+ &&&& HI & W& 25/50 & Rots 1975 \cr
\v

\+ NGC 3034 (M82)  Ir  80-90 &&& 3.25 & CO 2-1 & I30m &13 
& Sofue et al 1992 \cr
\+ &&&& HI & V & 18 & Yun et al. 1993\cr
\v

\+ NGC 3079 Sc & & $\sim90$&15.6 & \dotfill&\dotfill&\dotfill&Paper I\cr
\v

\+ NGC 3198 SBc && 70 &9.1 & HI &W & $25\times35$ & Bosma 1981a \cr
\+ &&&& HI &W & $25\times35$  &Begeman  1989 \cr 
\+ &&&& Opt & & &Hunter, et al. 1986 \cr
\v

\+NGC 3521     Sbc &&    75& 8.9 &HI &W& $74\times53$
& Casertano, van Gorkom 1991 \cr
\+&&&& CO &N45m&	15& Sofue et al 1997\cr
\v
		    
\+NGC 3628 Sb(Ir)  $>86$ &&& 6.7& CO &NMA &3.9& Irwin \& Sofue 1995\cr
\+ &&&& HI &V & 15&   Wilding et al 1993 \cr
\v

\+ NGC 4258 Sbc && 67&   6.6  &HI &W &40  & van Albada \& Shane 1976\cr
\+&&&&CO&N45m &15 & Sofue et al 1989 \cr   
\v

\+NGC 4303  Sc &&27&8.1& CO & N45m & 15 & Nishiyama 1995 \cr
\+ &&&& CO & NMA & 4 & Sofue et al. 1997 \cr
\+&&&&HI & V & 45 & Guhathakurta et al 1988 \cr
\v

\+ NGC 4321 (M100) Sc   27&&&15 &CO&N45m &15&   Nishiyama 1995 \cr
\+ &&&& CO &NMA & 	4 & Sakamosto 1995 \cr
\+ &&&& HI &V && Knapen et al 1993 \cr
\v

\+ NGC 4565 Sb & &86    &10.2 &\dotfill&\dotfill&\dotfill&Paper I\cr 
\v

\+NGC 4569 Sab & &63 &8.2 & CO & N45m & 15 & Nishiyama 1995 \cr
\+ &	&&& CO &NMA &$9.9\times4.8$ & Sofue et al 1997\cr
\+	&&&& HI &V  &20&  Guhathakurta \& van Gorkom \cr
\v

\+ NGC 4631  Sc Ir &  & 84   & 5.2&HI &W &$48\times89$ &Weliachev et al 1978\cr
\+&&&&CO&N45m &15 & Sofue et al.  1989, 1997\cr
\+&&&&CO(2-1)&I30m &13 & Sofue et al.  1990\cr
\v

\+ NGC 4736  Sab &&       35&5.1& CO&N45m&15 & Nishiyama 1995 \cr
\+&&&			 & HI &W& $25\times38$& Bosma et al 1977 \cr
\v

\+ NGC 4945  Sc I && 78& 6.7 & CO(2-1) &S15m & 24  &Dahlem et al 1993 \cr
\+ &&&& CO(1-0) &S15m&43 & ibid \cr
\+ &&&& HI &TE  &&Ables et al. 1987\cr
\v
 
\+ NGC 5033 Sc & &       62 &14& CO&N45m& 15&  Nishiyama 1995 \cr
\+ &&&& HI &W & $51\times85$ & Bosma 1981a \cr
\v

\+ NGC 5055 Sbc &&       55&8 &HI &W&    $49\times73$& Bosma 1981a \cr
\+&&&&	CO&N45m& 15& Sofue et al 1997\cr 
\v

\+ NGC 5194  (M51) Sc 20 &&&9.6  &\dotfill&\dotfill&\dotfill&Paper I\cr 
\v

\+ NGC 5236 (M83) SBc  24 &&&8.9& CO &N45m &15 &
Handa et al 1978, Nishiyama 1995 \cr
\+ &&&& HI &W & 40 & Bosma et al. 1981 \cr
\v

\+ NGC 5457 (M101)  Sc  18 &&&     7.2&HI &W &$24\times30$ &  Bosma et al. 1981 \cr 
\+ &&&   &CO &F14m &50& Kenney et al 1991 \cr 
\v

\+ NGC 5907 Sc && 88& 11.6&\dotfill &\dotfill&\dotfill& Paper I\cr
 \v

\+ NGC 6946 Sc  & & 30   & 5.5 &\dotfill&\dotfill&\dotfill& Paper I\cr  
\v

\+ NGC 6951 Sbc & &48 && CO &N45m&15&  Nishiyama 1995 \cr
\+ && && Opt  &&&  Marquez, Moles 1993 \cr
\v

\+NGC 7331      Sbc &&   75&     14& HI &W & $25\times45$ & Bosma 1981a \cr
\+ &&&& CO&N45m&15& Sofue et al 1997\cr
\v

\+ Milky Way Sb && 90 & 0 & CO+HI & & & Clemens 1995 \cr
\+  ~~~~~~(outside solar circle) &&&& HI &&& Honma \& Sofue 1996a,b \cr

\v

\hrule
\v

* Distances have been taken from the references in the
same row. For the original data, refer to the literature.

$\dagger$
(CO-line observations) 
 N45m = NRO 45-m telescope; 
I30m = IRAM 30-m telescope; 
S15m = SEST 15m telescope;  
F14m = FCRAO 14-m telescope; 
K12m = NRAO Kitt-Peak 12-m telescope;
NMA = Nobeyama Millimeter Array.
 (HI-line observations) E100m = Effelsberg 100-m telescope; 
CI = Cambridge Interferometer; 
W = WSRT; 
V = VLA; 
TE = Two-Element Synthesis telescope. 
(Optical observations) Opt = Optical line (e.g., H$\alpha$) observations.

\endpage

\r Figure Captions

\v

\r Fig. 1: Examples of position-velocity diagrams in the CO
line emission for some galaxies, and traced rotation curves,
which are smoothly connected with HI rotation curves.
Crosses indicate typical errors (vertical) and angular
resolution (horizontal).

\v
\r Fig. 2: Inner-to-out rotation curves obtained
by combining CO and HI rotation curves of galaxies. Some curves
have been obtained by using optical data.
Typical angular resolutions and errors are indicated by crosses
at several representative points for each object.
The horizontal bars indicate the  angular resolutions,  and the
vertical bars give errors in the rotation velocity estimates.
Dashed vertical lines indicate the borders of CO and HI
rotation curves. Borders of regions with different data
of different angular resolutions are also marked by
dashed lines. Optical images from the STScI Digital
Sky Survey are shown with the major axes and angular scales
superposed.

\bye